\documentclass[prb,aps,epsfig,twocolumn,showpacs,float]{revtex4}
\usepackage{graphicx}
\usepackage{amssymb}
\usepackage{amsmath}
\begin{document}

\draft

\title{Scaling of intrinsic Gilbert damping with spin-orbital coupling strength}

\author{P. He$^{\mathrm{1,4}}$, X. Ma$^{\mathrm{2}}$, J. W. Zhang$^{\mathrm{4}}$, H. B. Zhao$^{\mathrm{2,3}}$, G. L\"{u}pke$^{\mathrm{2}}$, Z. Shi$^{\mathrm{4}}$, and S. M. Zhou$^{\mathrm{1, 4}}$}

\address{$^{\mathrm{1}}$Surface Physics State Laboratory and Department of Physics, Fudan University, Shanghai 200433, China}
\address{$^{\mathrm{2}}$Department of Applied Science, College of William and Mary, Williamsburg, Virginia 23185}
\address{$^{\mathrm{3}}$Key Laboratory of Micro and Nano Photonic Structures (Ministry of Education) and Department of Optical Science and Engineering, Fudan University, Shanghai 200433, China}
\address{$^{\mathrm{4}}$Shanghai Key Laboratory of Special Artificial Microstructure Materials and Technology $\&$ Physics Department, Tongji University, Shanghai 200092, China}

\date{\today}

\begin{abstract} We have experimentally and theoretically investigated the dependence of the intrinsic Gilbert damping parameter $\alpha_0$ on the spin-orbital coupling strength $\xi$ by using L1$_{\mathrm{0}}$~ordered FePd$_{\mathrm{1-x}}$Pt$_{\mathrm{x}}$ ternary alloy films with perpendicular magnetic anisotropy. With the time-resolved magneto-optical Kerr effect, $\alpha_0$ is found to increase by more than a factor of ten when $x$ varies from 0 to 1.0. Since changes of other leading parameters are found to be neglected, the $\alpha_0$ has for the first time been proven to be proportional to $\xi^2$.

\end{abstract}

\vspace{0.5 cm}

\pacs{75.78.Jp; 75.50.Vv; 75.70.Tj; 76.50.+g} \maketitle

\indent Magnetization dynamics has currently become one of the most popular topic in modern magnetism due to its crucial importance in information storage. Real space trajectory of magnetization processional switching triggered by magnetic field pulses, fs laser pulses, and spin-polarized current~\cite{1a,1b,2a,2b,3b,3c}, can be well described by the phenomenological Landau-Lifshitz-Gilbert (LLG) equation that incorporates the Gilbert damping term~\cite{4a} which controls the dissipation of magnetic energy towards the thermal bath. The time interval from the non-equilibrium magnetization to the equilibrium state is governed by the Gilbert parameter $\alpha$. It has very recently been shown that the laser-induced ultrafast demagnetization is also controlled by the $\alpha$~\cite{3a}.\\
\indent The intrinsic Gilbert damping $\alpha_0$ has been extensively studied in theory ~\cite{7a,7b,7c,8a,9a,9b,9c}, and in general believed to arise from the spin orbital coupling (SOC). In the SOC torque-correlation model proposed by Kambersk\'{y}, contributions of intraband and interband transitions are thought to play a dominant role in the $\alpha_0$ at low and high temperatures $T$ and are predicted to be proportional to $\xi^{3}$ ($\xi$=the SOC strength) and $\xi^{2}$, respectively~\cite{7b,9b}. Up to date, however, no experiments have been reported to demonstrate the quantitative relationship between $\alpha_0$ and $\xi$ although many experimental attempts have been made to study the $\alpha_0$ in various metallic and alloy films~\cite{11a,12a,13a,14a,15a,18a,19a,19b}. It is hard to rule out effects other than the SOC because $\alpha_0$ is also strongly related to parameters such as the electron scattering time and density of state $D(E_{F})$ at Fermi surface $E_{F}$~\cite{13a,18a,22b} which in turn change among various metals and alloys. In order to rigorously address the $\xi$ dependence of $\alpha_0$ in experiments, it is therefore essential to find magnetic alloys in which the $\xi$ can be solely adjusted while other parameters almost keep fixed. \\
\indent In this Letter, we elucidate the $\xi$ dependence of $\alpha_0$ by using L1$_{\mathrm{0}}$~FePd$_{\mathrm{1-x}}$Pt$_{\mathrm{x}}$ (=FePdPt) ternary alloy films. Here, only $\xi$ is modulated artificially by the Pt/Pd concentration ratio because heavier atoms are expected to have a larger $\xi$~\cite{27a,27b,27c} and parameters other than $\xi$ are theoretically shown to be almost fixed. Experimental results have shown that $\alpha_0$ is proportional to $\xi^2$. It is therefore the first time to have given the experimental evidence of the $\xi^2$ scaling law. This work will also facilitate exploration of new magnetic alloys with reasonably large perpendicular magnetic anisotropy (PMA) and low $\alpha$.\\
\indent L1$_{\mathrm{0}}$ FePdPt ternary alloy films with $0\leq x \leq 1.0$
were deposited on single crystal MgO (001)
substrates by magnetron sputtering. The FePdPt composite target was formed by putting small Pt and Pd pieces
on an Fe target. During deposition, the substrates were kept at 500 $\mathrm{^{\circ}C}$.
After deposition, the samples were annealed in situ at the same
temperature for 2 hours. The base pressure of the deposition
system was $1\times$10$^{\mathrm{-5}}$ Pa and the Ar pressure was
0.35 Pa. Film thickness was determined by X-ray reflectivity (XRR) to be $12 \pm 1$ nm. In order to measure the Gilbert damping parameter $\alpha$~\cite{28a,29a}, time-resolved magneto-optical Kerr effect (TRMOKE) measurements were performed in a
pump-probe setup using a pulsed Ti:sapphire laser in the wavelength of
400 nm (800 nm) for pump (probe) pulses with a pulse duration of
200 fs and a repetition rate of 250 kHz. An intense pump pulse
beam with a fluence of 0.16 mJ/cm$^{\mathrm{2}}$ was normally
incident to excite the sample, and the transient Kerr signal was
detected by a probe pulse beam which is
time-delayed with respect to the pump. The intensity ratio of the
pump to probe pulses was set to be about 6:1, and their respective
focused spot diameters were 1 mm and 0.7 mm. A variable magnetic
field $H$ up to 5 T was applied at an angle of 45 degrees with respect
to the film normal using a superconducting magnet. TRMOKE measurements were performed at 200 K and other measurements were performed at room temperature.\\
\begin{figure}
\begin{center}
\resizebox*{3 in}{!}{\includegraphics*{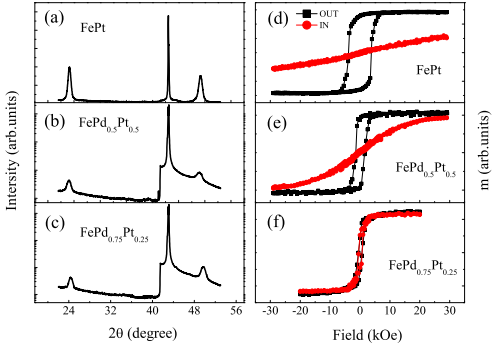}}
\caption{XRD patterns(a, b, c), out-of-plane and
in-plane hysteresis loops (d,e,f)for L1$_{\mathrm{0}}$
FePd$_{\mathrm{1-x}}$Pt$_{\mathrm{x}}$ films with $x=1$ (a,d), $x=0.5$ (b,e) and $x=0.25$ (c,f).} \label{Fig1}
\end{center}
\end{figure}
\indent Microstructural analysis was accomplished with the aid of X-ray diffraction (XRD). Figures~\ref{Fig1}(a)-\ref{Fig1}(c) show
the XRD patterns for L1$_{\mathrm{0}}$ FePdPt films with $x=1$, $x=0.5$, and $x=0.25$, respectively. The films are of the L1$_{\mathrm{0}}$ ordered structure in the presence of (001) superlattice peak. The chemical ordering degree $S$ can be calculated with the intensity of the (001) and (002) peaks and found to be $0.7\pm0.1$ for all samples. Since no other diffraction peaks exist except for (001) and (002) ones, all samples are of L1$_{\mathrm{0}}$ single phase with \emph{c} axis perpendicular to the film plane. Here, $c=3.694 {\AA}$. Magnetization hysteresis loops were measured by vibrating sample magnetometer.
Figures~\ref{Fig1}(d)-\ref{Fig1}(f) display the corresponding out-of-plane and in-plane magnetization hysteresis loops. As shown in
Fig.\ref{Fig1}(d), for $x=1$ (L1$_{\mathrm{0}}$ FePt) the
out-of-plane hysteresis loop is almost square-shaped with coercivity
$H_{\mathrm{C}}=3.8$ kOe, indicating the establishment of high
PMA. With decreasing $x$, the $H_{\mathrm{C}}$ decreases. For $x=0.25$ in Fig.~\ref{Fig1}(f),
$H_{\mathrm{C}}$ approaches zero and
the out-of-plane and in-plane loops almost overlap with each other, indicating a weak PMA. Apparently, the PMA increases with increasing $x$. Similar phenomena have been reported elsewhere~\cite{27b,27c}.\\
\begin{figure}
\begin{center}
\resizebox*{3 in}{!}{\includegraphics*{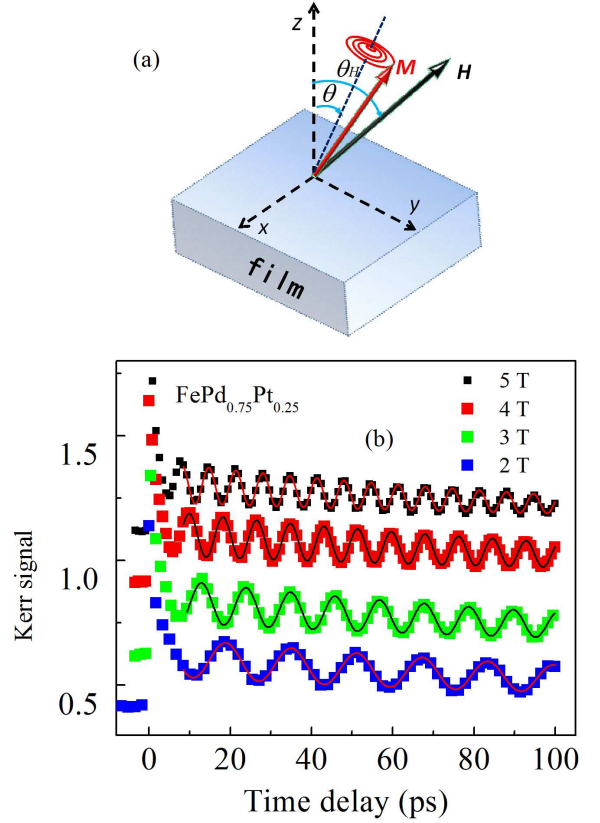}}
\caption{Schematic illustration of the TRMOKE geometry (a) and TRMOKE results for $x=0.25$ under various
magnetic fields (b). Here $\theta_{H}=45^{\circ}$. Curves are shifted for clarity. The
solid lines are fit results.}
\label{Fig2}
\end{center}
\end{figure}
\indent Figure \ref{Fig2}(b)
displays the typical TRMOKE results for L1$_{\mathrm{0}}$
FePdPt films with $x=0.25$ under $\theta_{\mathrm{H}}=45^{o}$ as shown in Fig.\ref{Fig2}(a). For the time delay longer than 5.0 ps, damped oscillatory Kerr signals are clearly seen due to the magnetization precession. The precession period becomes short significantly with increasing $H$. In order to extract the precession frequency, the Kerr signal was fitted by following
exponentially damped sine function, $a+b
\exp(-t/t_{0})+A\exp(-t/\tau)\sin(2\pi\emph{f}t+\varphi)$, where parameters $A$, $\tau$, $\emph{f}$ and $\varphi$ are the amplitude, relaxation time, frequency, and phase of damped magnetization
precession, respectively~\cite{30a}. Here, $a$, $b$, and $t_{0}$ correspond to the background signal owing to the slow recovery process. The experimental data are well fitted by the above equation, as shown in Fig.\ref{Fig2}(b).\\
\begin{figure}
\begin{center}
\resizebox*{3 in}{!}{\includegraphics*{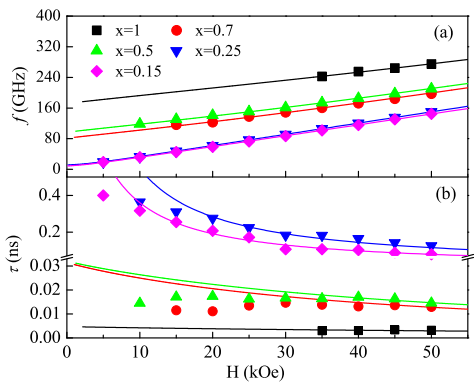}} \caption{Uniform
magnetization precession frequency \emph{f} (a) and relaxation
time $\tau$ (b) as a function of $H$ for
all samples studied here. Solid lines refer to fit results.}\label{Fig3}
\end{center}
\end{figure}
\indent Figure~\ref{Fig3}(a) shows that for all samples studied here,
the extracted precession frequency $f$
increases monotonically as $H$ increases. Moreover, $f$ shows
an increasing tendency with increasing $x$ at fixed $H$. For $x=1$
(L1$_{\mathrm{0}}$ FePt), $f$ is in a very high frequency range of
180-260 GHz due to the high PMA. Figure \ref{Fig3}(b) shows that the relaxation time $\tau$ displays a decreasing trend with
increasing $H$. Moreover, $\tau$ increases by two orders of
magnitude when Pd atoms are replaced by Pt ones. In particular, we
observed the short relaxation time $\tau=3$ ps for $x=1$ (L1$_{\mathrm{0}}$ FePt). When the oscillation period is longer than the relaxation time
for low $H$ the precession cannot be excited for $x=1$~\cite{30c}.\\
\indent With $\alpha\ll 1.0$, one can obtain the following dispersion equation, $2\pi\emph{f}=\gamma\sqrt{H_{1}H_{2}}$, where $H_{1}=H\cos(\theta_{H}-\theta)+H_{K}\cos2\theta$ and $H_{2}=H\cos(\theta_{H}-\theta)+H_{K}\cos^{2}\theta$, where $H_{K}=2K_{U}/M_{S}-4\pi M_{S}$ with uniaxial anisotropy constant $K_{\mathrm{U}}$. The equilibrium magnetization angle $\theta$ is calculated from the following equation
$\sin2\theta=(2H/H_{K})\sin(\theta_{H}-\theta)$, which is derived by taking the minimum of the total free energy. The measured $H$ dependence of \emph{f} can be well fitted, as shown in Fig.\ref{Fig3}(a). With the measured $M_{S}$ of 1100 emu/cm$^{3}$, the $K_{U}$ can be calculated. The $g$ factor is equal to 2.16 for $x=1$, 0.7, and 0.5, and to 2.10 and 2.03 for $x=0.25$ and 0.15, respectively. A small fraction of the orbital angular momentum is therefore restored by the SOC~\cite{7b} and close to results reported elsewhere~\cite{30b}.\\
\indent The measured $H$ dependence of $\tau$ can be well fitted by $\tau=2/\alpha\gamma(H_{1}+H_{2})$ with the fitted values of $g$ and $H_{\mathrm{K}}$ for $\alpha \ll 1.0$. Here, the Gilbert damping $\alpha$ is an adjustable parameter. As shown in Fig.\ref{Fig3}(b), the experimental and
fitted data coincide with each other at high $H$ and exhibit significant deviation from each other at low $H$. It is therefore illustrated that the extrinsic magnetic relaxation contributes to the $\alpha$ at low $H$ and becomes weak at high $H$. This is because
the extrinsic magnetic relaxation may arise from the inhomogeneous PMA distribution and the interfacial effect and is greatly suppressed
under high $H$~\cite{25b,31a,25c}. The intrinsic $\alpha_{0}$ therefore plays a dominant role at high $H$, that is to say, $\alpha_{0}$ is fitted in Fig.~\ref{Fig3}(b). \\
\begin{figure}
\begin{center}
\resizebox*{3 in}{!}{\includegraphics*{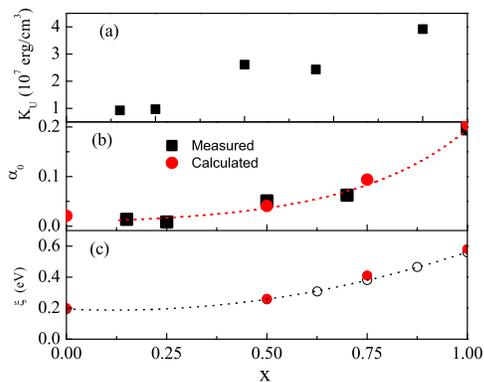}} \caption{Measured $K_{\mathrm{U}}$ (a), measured (solid box) and calculated (solid circles) $\alpha_{0}$ (b), $\xi$ calculated in this work (solid circles) and elsewhere~\cite{35a} (open ones) (c) as a function of $x$. The lines serve as a visual guide in (b) and refer to the fit results in (c).}\label{Fig4}
\end{center}
\end{figure}
\indent To determine the SOC strength $\xi$ and intrinsic damping parameter $\alpha_0$ in L1$_{\mathrm{0}}$~ordered FePd$_{\mathrm{1-x}}$Pt$_{\mathrm{x}}$ ternary alloys, we perform spin dependent first principles calculations based on linear muffin-tin orbital density functional theorem, where the lattice constants are $a=3.86\AA$ and $c=3.79\AA$ for L1$_{\mathrm{0}}$ ordered FePt. The $D(E_{F})$ is 2.55, 2.47, 2.43, and 2.39 per atom per eV for $x$ varying from 0, 0.5, 0.75, to 1.0, respectively. The $\alpha_0$ was achieved by using spin-orbital torque-correlation model based on spin dependent electron band structures obtained above\cite{7a,9a}.\\
\indent It is significant to compare variations of the PMA and $\alpha_{0}$. Figures~\ref{Fig4}(a)~$\&$~\ref{Fig4}(b) show the $K_{U}$ and $\alpha_{0}$ both decrease with decreasing $x$. Similar variation trends of $K_{U}$ and $\alpha_{0}$ have been observed for perpendicularly magnetized Pt/Co/Pt multilayers~\cite{30a}. When the $\xi$ is smaller than the exchange splitting, the magnetic anisotropy is thought to come from the second order energy correction of the SOC in the perturbation treatment and is roughly proportional to both the $\xi$ and the orbital angular momentum. The orbital momentum in 3\emph{d} magnetic metallic films restored by the SOC is also proportional to the $\xi$ and the PMA therefore is proportional to $\xi^{\mathrm{2}}/W$ with the bandwidth of 3\emph{d} electrons $W$~\cite{32a}. Since the $W$ does not change much with $x$, the enhanced PMA at high $x$ is attributed to a larger $\xi$ of Pt atoms compared with that of Pd atoms~\cite{27a,34a}. Our calculations show $\xi$ change from $0.20$, $0.26$, $0.41$ to $0.58$ (eV) when $x$ varying from $0$, $0.5$, $0.75$, to $1.0$, as shown in Fig.~\ref{Fig4}(c). This is because the $\xi$ is 0.6, 0.20, and 0.06 (eV) for Pt, Pd, and Fe atoms, respectively~\cite{27a,35a} and the effect of Fe atoms is negligible compared with those of Pd and Pt atoms. The present results of $\xi$ are in good agreement with previous \emph{ab initio} calculations~\cite{35a}. Apparently, the PMA behavior arises from the increase of $\xi$ at high $x$. As shown in Fig.~\ref{Fig4}(b), measured and calculated results of $\alpha_{0}$ are in a good agreement.
Since the lattice constant, $D(E_{F})$, the Curie temperature, the gyromagnetic ratio, and the averaged spin are experimentally and theoretically shown to almost not change with $x$, the enhanced $\alpha_{0}$ is mainly attributed to the $\xi$ increase with increasing $x$. Figure~\ref{Fig5} shows that the $\alpha_{0}$ is approximately proportional to $\xi^2$, where the $\xi$ values at other $x$ are exploited from the fitted curve in Fig.~\ref{Fig4}(c). Since for the present L1$_{\mathrm{0}}$~ordered FePd$_{\mathrm{1-x}}$Pt$_{\mathrm{x}}$ ternary alloy films only $\xi$ is tuned with $x$, the  present work has rigorously proven the theoretical prediction about the $\xi^2$ scaling of $\alpha_{0}$~\cite{7a}. It is indicated that the $\alpha_0$ at 200 K is mainly contributed by the interband contribution\cite{7b,7c,9b}. The electronic-scattering-based model of ferromagnetic relaxation is therefore proved to be applicable for the $\alpha_{0}$ in L1$_{\mathrm{0}}$ FePdPt ternary alloys~\cite{7a}. In order to further verify the $\xi^3$ dependence of $\alpha_{0}$~\cite{9b}, measurements of magnetization precession at low temperatures need to be accomplished.\\
\begin{figure}
\begin{center}
\resizebox*{3 in}{!}{\includegraphics*{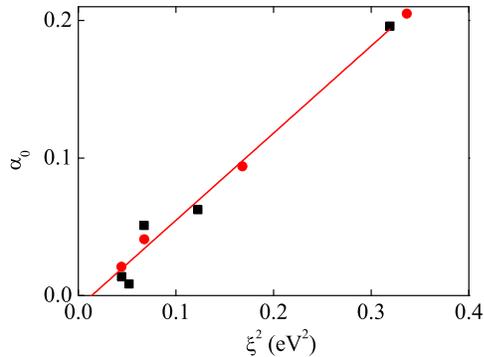}} \caption{The measured (solid square) and calculated (solid circles) $\alpha_{0}$ versus $\xi^2$ as a function of $x$. The dashed curve refers to the linear fit results.}\label{Fig5}
\end{center}
\end{figure}
\indent In summary, we have investigated the magnetization
dynamics in L1$_{\mathrm{0}}$
FePdPt ternary alloy films using TRMOKE. The intrinsic $\alpha_{0}$ can be continuously tuned, showing a decrease
with increasing Pd content due to smaller $\xi$ compared with that of Pt atoms. In particular, the $\xi^2$ dependence of $\alpha_{0}$ has been rigorously demonstrated in experiments. The experimental results deepen
the understanding the mechanism of $\alpha_{0}$ in magnetic metallic materials and provide a new clue to explore ideal ferromagnets with reasonably low $\alpha_{0}$ and high PMA as storage media for the next generation microwave-assisted magnetic recording.\\
\indent Acknowledgements This work was supported by the MSTC under grant
No. 2009CB929201, (US) DOE grant No. DE-FG02-04ER46127, NSFC under Grant
Nos. 60908005, 51171129 and 10974032, and Shanghai PuJiang Program (10PJ004). \\


\begin{thebibliography}{27}

\bibitem{1a} Th. Gerrits, H. A. M. van den Berg, J. Hohlfeld, L. B\"{a}r, and Th. Rasing, Nature (London) \textbf{418}, 509 (2002).
\bibitem{1b} H.W. Schumacher, C. Chappert, R. C. Sousa, P. P. Freitas, and J. Miltat, Phys. Rev. Lett. \textbf{90}, 017204 (2003).

\bibitem{2a} S. I. Kiselev, J. C. Sankey, I.N. Krivorotov, N.C. Emley, R. J. Schoelkopf, R. A. Buhrman, and D. C. Ralph, Nature
(London) \textbf{425}, 380 (2003).

\bibitem{2b}S. Kaka, M. R. Pufall, W. H. Rippard, T. J. Silva, S. E. Russek, and J. A. Katine, Nature (London) \textbf{437}, 389
(2005)

\bibitem{3b} E. Beaurepaire, J. C. Merle, A. Daunois, and J.Y. Bigot, Phys. Rev. Lett. \textbf{76}, 4250 (1996).

\bibitem{3c} J. Hohlfeld, E. Matthias, R. Knorren, and K. H. Bennemann, Phys. Rev. Lett. \textbf{78}, 4861 (1997).

\bibitem{4a}T. L. Gilbert, Phys. Rev. \textbf{100}, 1243(1955); L. D. Landau, E. M. Lifshitz, and L. P. Pitaevski, \emph{Statistical Physics}, Part 2
(Pergamon, Oxford,1980), 3rd ed.

\bibitem{3a}B. Koopmans, J. J. M. Ruigrok, F. Dalla Longa, and W. J. M. de Jonge, Phys. Rev. Lett. \textbf{95}, 267207(2005)

\bibitem{7a}V. Kambersk$\acute{y}$, Can. J. Phys. \textbf{48}, 2906 (1970)

\bibitem{7b} V. Kambersk$\acute{y}$, Czech. J. Phys., Sect. B \textbf{26}, 1366 (1976)

\bibitem{8a} J. Kunes and V. Kambersky, Phys. Rev. B \textbf{65}, 212411 (2002)

\bibitem{7c}V. Kambersk$\acute{y}$, Phys. Rev. B \textbf{76}, 134416(2007)


\bibitem{9a} K. Gilmore, Y. U. Idzerda, and M. D. Stiles, Phys. Rev. Lett. \textbf{99}, 027204(2007)

\bibitem{9b}K. Gilmore, Y. U. Idzerda, and M. D. Stiles, J. Appl. Phys.~\textbf{103}, 07D303(2008)

\bibitem{9c} H. Ebert, S. Mankovsky, D. K\"{o}dderitzsch, and P. J. Kelly, Phys. Rev. Lett. \textbf{107}, 066603 (2011)


\bibitem{11a}C. E. Patton, Z. Frait, and C. H. Wilts, J. Appl. Phys. \textbf{46}, 5002(1975)

\bibitem{12a}S. Mizukami, Y. Ando, and T. Miyazaki, J. Magn. Magn. Mater. \textbf{226-230}, 1640(2001)

\bibitem{19a} S. Ingvarssona, G. Xiao, S. S. P. Parkin, and R. H. Koch, Appl. Phys. Lett. \textbf{85}, 4995(2004)

\bibitem{14a}Y. Guan and W. E. Bailey, J. Appl. Phys. \textbf{101}, 09D104(2007)

\bibitem{15a} C. Scheck, L. Cheng, I. Barsukov, and Z. Frait \emph{et al}, Phys. Rev. Lett. \textbf{98}, 117601(2007)

\bibitem{18a}J. O. Rantschler, R. D. McMichael, A. Castillo, and A. J. Shapiro \emph{et al}, J. Appl. Phys. \textbf{101}, 033911(2007)

\bibitem{19b}G. Woltersdorf, M. Kiessling, G. Meyer, J.-U. Thiele, and C. H. Back, Phys. Rev. Lett. \textbf{102}, 257602(2009)

\bibitem{13a}A. A. Starikov, P. J. Kelly, A. Brataas, and Y. Tserkovnyak \emph{et al}, Phys. Rev. Lett. \textbf{105}, 236601(2010)

\bibitem{22b}
S. Mizukami, D. Watanabe, M. Oogane,  and Y. Ando \emph{et al}, J. Appl. Phys. \textbf{105}, 07D306 (2009)





\bibitem{28a}W. K. Hiebert, A. Stankiewicz, and M. R. Freeman, Phys. Rev. Lett. \textbf{79}, 1134(1997)

\bibitem{29a} M. van Kampen, C. Jozsa, J. T. Kohlhepp, and P. LeClair \emph{et al}, Phys.
Rev. Lett. \textbf{88}, 227201(2002)

\bibitem{27a}
K. M. Seemann, Y. Mokrousov, A. Aziz, and J. Miguel \emph{et al}, Phys. Rev. Lett. \textbf{104}, 076402(2010).

\bibitem{27b}
S. Jeong et al., J. Appl. Phys. \textbf{91}, 8813(2002)

\bibitem{27c}
G. J. Chen et al., Surf. Coat. Technol. \textbf{202}, 937(2007)

\bibitem{30a}
S. Mizukami, E. P. Sajitha, F. Wu, and D. Watanabe~\emph{et al}, Appl. Phys. Lett. \textbf{96},
152502(2010)

\bibitem{30c}
J. W. Kim, H. S. Song, J. W. Jeong, K. D. Lee~\emph{et al}, Appl. Phys. Lett. \textbf{98}, 092509(2011)

\bibitem{30b}
I. V. Solovyev, P. H. Dederichs, and I. Mertig, Phys. Rev. B \textbf{52}, 13419(1995)

\bibitem{31a}J. Walowski, M. Djordjevic-Kaufmann, B. Lenk, and C. Hamann \emph{et al}, J. Phys. D: Appl. Phys. \textbf{41}, 164016(2008)

\bibitem{25b}R. Urban, G. Woltersdorf, and B. Heinrich, Phys. Rev. Lett. \textbf{87}, 217204(2001)

\bibitem{25c}Y. Tserkovnyak, A. Brataas, and G. E. W. Bauer, Phys. Rev. Lett. \textbf{88}, 117601(2002)





\bibitem{32a} P. Bruno, Phys. Rev. B~\textbf{39}, 865(1989)


\bibitem{34a}
J. Friedel, in \emph{The Physics of Metals}, edited by J. M. Ziman (Cambridge Univ. Press, Cambridge, 1969)

\bibitem{35a} P. He, L. Ma, Z. Shi, G. Y. Guo, and S. M. Zhou,~\emph{Chemical Composition Tuning of the Anomalous Hall Effect in Isoelectronic L1(0) FePd$_{1-x}$Pt$_{x}$ Alloy Films}, arXiv:1112.0834v1



\end{thebibliography}
\end{document}